\documentclass{moriond}

\def\Journal#1#2#3#4{{#1} {\bf #2}, #3 (#4)}

\def\PRL{\em Phys. Rev. Lett.}

\def\PRA{{\em Phys. Rev.} A}
\def\GRR{\em Gen. Relat. Gravit.}
\def\NAT{\em Nature}
\def\JPB{{\em J. Phys.} B}
\def\POP{\em Phys. Plasmas}
\def\APL{\em Appl. Phys. Lett.}
\def\NIR{{\em Nucl. Instrum. Methods Phys. Res., Sect.} B}
\def\NJP{\em New J. Phys.}
\def\JMP{\em Int. J. Mod. Phys. Conf. Ser.}
\def\JPS{\em JPS Conf. Proc.}
\def\RMF{\em Ref. Mod. Phys.}
\def\HII{\em Hyperfine Interact}

\def\be{\begin{equation}}
\def\ee{\end{equation}}
\def\bea{\begin{eqnarray}}
\def\eea{\end{eqnarray}}

\newcommand{\Photo}{}

\begin{document}
\vspace*{4cm}
\title{Status of the GBAR experiment at CERN}

\author{Barbara Latacz on behalf of the GBAR collaboration}

\address{IRFU, CEA Paris-Saclay\\
F-91191 Gif sur Yvette Cedex, France}

\maketitle\abstracts{
The GBAR experiment aims at measuring the free fall of antihydrogen atoms in the terrestrial gravitational field. It is located at CERN in the AD area. It is the first experiment which has been connected to the ELENA low energy antiproton ring that started commissioning in the summer of 2018. First tests with antiproton and positron beams were performed in summer and fall of 2018. The status and plans of the experiment are described in this document.}

\section{Introduction}

The main goal of the {\bf{GBAR}} collaboration is to measure the {\bf{G}}ravitational {\bf{B}}ehaviour of {\bf{A}}ntihydrogen at {\bf{R}}est. It is done by measuring the classical free fall of neutral antihydrogen, which is a direct test of the weak equivalence principle for antimatter. The experiment is based on an idea of J. Walz and T. H\"ansch \cite{Ch2-GBAR_principle}, which was brought to life by P. P\'erez \cite{Ch2-Patrice_first_talk,Ch1-GBAR-proposal}.

The first step of the experiment is to produce an antihydrogen ion $\bar{H}^{+}$ and catch it in a Paul trap, where it can be cooled to $\mu$K temperature using ground state Raman sideband sympathetic cooling. The $\mu$K temperature corresponds to a particle velocity in the order of 1 m/s. Once such velocity is reached, the antihydrogen ion can be neutralised and starts to fall. Due to the low temperature, a fall from 10 cm height corresponds to about 0.14 s time of flight. Such a long free fall allows achieving 37 \% error on the measurement of the gravitational acceleration $\bar{g}$ for antihydrogen after only one event. With this method, the aim is to reach 1 \% precision with about 1500 events \cite{Ch1-GBAR-proposal}.

The second goal of the experiment is to reach $10^{-5}-10^{-6}$ precision in the measurement of the gravitational quantum states of cold antihydrogen \cite{Ch2-Neswitch,Ch2-Nesvitch3}. This method is inspired by a similar experiment performed for cold neutrons \cite{Ch2-Neswitch2}.

\section{Production of antihydrogen ions}

The first part of the GBAR experiment is the production of the antihydrogen ion. 
It is made in the following reactions:

\begin{equation}Ps^{*}+\bar{p}\rightarrow\bar{H}^{*}+e^{-},\label{one}\end{equation}

\begin{equation}Ps^{*}+\bar{H}\rightarrow\bar{H}^{+}+e^{-}.\label{two}\end{equation} 

\noindent The ``$*$'' next to an atom symbol indicates that it can be either in the ground state or in an excited state. 

\begin{figure}
	\centering
	\includegraphics[width=0.8\textwidth]{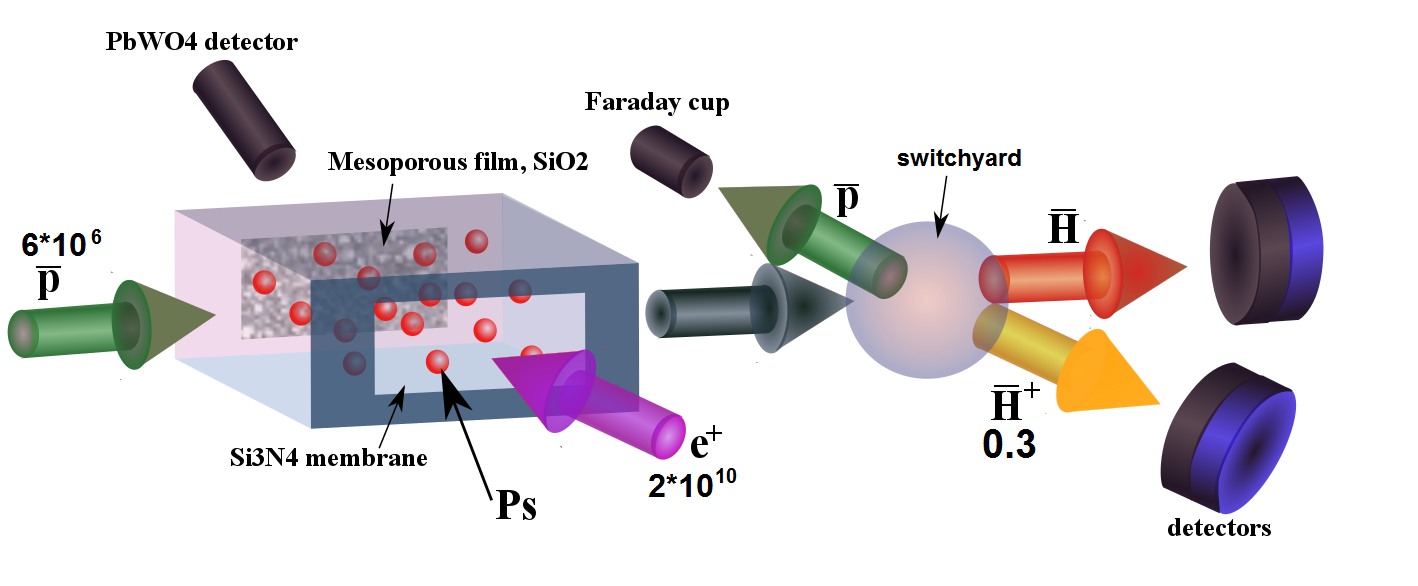}
	\caption{The general scheme of the antihydrogen ion production setup.}
	\label{Ch19-scheme}
\end{figure}

The general scheme of the antihydrogen ion production part of the GBAR experiment is shown in Figure \ref{Ch19-scheme}. The antihydrogen atom $\bar{H}$ and ion $\bar{H^{+}}$ production takes place in a cavity. Due to relatively low positronium energy in comparison to antiproton energy, the heaviest reaction products (antihydrogen atoms and ions) combined with the $\bar{p}$ beam are guided to the switchyard where they are separated in an electrostatic field (Figure \ref{Ch19-scheme}).

The production of a few $\bar{H}^{+}$ ions in one beam crossing requires about $10^{7}$ $\bar{p}$/bunch and a few $10^{11}$ Ps/cm$^{-3}$ positronium density inside a cavity, which is produced with a beam containing $5\cdot 10^{10}$ positrons per bunch. 

The $\bar{H}^{+}$ production part of the experiment was installed at CERN in 2018. Nowadays, the experiment is being commissioned with positrons and protons, which allow to perform the symmetric reactions of hydrogen atom and ion formation. An optimisation of the ion production with matter will help to be fully prepared for the next antiproton beam time in 2021.

\subsection{ELENA and antiproton deceleration}

GBAR is the only experiment that has received the antiproton beam from the new CERN antiproton decelerator - ELENA \cite{Ch1-ELENA}. The first ELENA beam reached the experiment in June 2018 and was tested until November 2018. After many improvements performed by both ELENA and GBAR teams, we can report that ELENA reached the nominal beam parameters at the extraction point. This indicates that in spring 2021 the decelerator should be ready to work with all experiments.

The ELENA beam has a 100 keV energy and an intensity reaching $5\cdot10^{6}$ $\bar{p}$/bunch. This beam is later decelerated to 1-10 keV energy \cite{Ch2-decelerator} in order to obtain the highest possible $\bar{H^{+}}$ formation rate \cite{Ch2-cross1,Ch2-cross2}. From next year, a high field Penning-Malmberg trap will be added to the system in order to trap antiprotons. This will allow obtaining better beam parameters.

\begin{figure}
	\centering
	\includegraphics[width=0.5\textwidth]{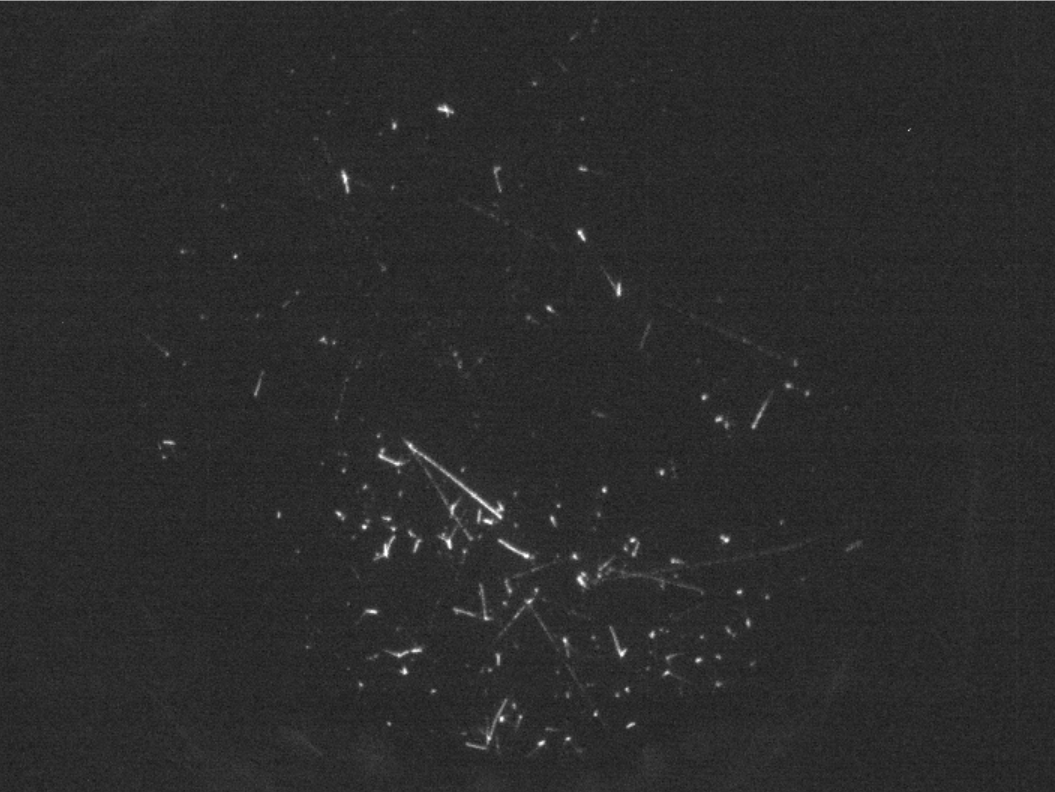}
	\caption{An MCP image of single 10 keV antiprotons.}
	\label{MCP_antiprotons}
\end{figure}

Due to technical difficulties on both ELENA and GBAR sides, it was possible to test the deceleration system only partially. Additionally, the first tests of the detection system were performed. Example image of single 10 keV antiprotons registered with the MCP detector that will be used for $\bar{H}$ detection is shown in Figure \ref{MCP_antiprotons}.

\subsection{The slow positron source} 

The GBAR positron source is based on a 9 MeV linear electron accelerator. The relatively low energy was chosen to avoid activation of the environment with gamma radiation. The electron beam is incident on a tungsten target where positrons are created from Bremsstrahlung radiation (gammas) through the pair creation process. Some of the created positrons undergo a~further diffusion in the tungsten moderator reducing their energy to about 3 eV.
The created particles are re-accelerated to about 53 eV energy and are adiabatically transported to the next stage of the experiment.
Presently, the measured positron flux is equal to\break  $N_{e^{+}/1s}=6.02 \pm  0.45 (stat) \pm 1.55 (sys)\cdot 10^{7}~$e$^+$/s, which is a few times higher than intensities reached with radioactive sources.

\subsection{Positron trapping}

The goal of the GBAR experiment is to obtain more than $10^{10}$ e$^{+}$/pulse. In order to achieve it, a two-stage trapping system is used.

The first device \cite{Ch2-gbartrap} is based on a Surko buffer gas trap \cite{CH2-posi-surko}. The main goal of this 40 mT magnetic field trap is to cool positrons and reduce their radial velocity \cite{Ch2-surko1,Ch2-surko2}, so that they can be trapped for long times and even accumulated. Presently, the trapping efficiency is equal to 5~\%.
The second device is a Penning Malmberg trap with a high magnetic field of 5 T. In this trap, $10^{10}$~e$^{+}$ are going to be stored in the form of a non-neutral plasma \cite{Ch2-plasma2}.  At this time the measured positron flux after the trap is equal to $10^{8}$ e$^{+}$/pulse. Currently, the trap is using a synchrotron cooling technique, but later the goal is to use an electron cooling technique.

\subsection{Positronium formation}

The production of antihydrogen ions requires a cold (energy $< 0.1$ eV) dense positronium target. It is produced by positron implantation on a mesoporous silica target, for which the positron to ortho-positronium conversion probability is equal to 30 \%. The Ps formation has been tested before by collaborators \cite{Ch2-Laszlo1,Ch2-Laszlo2,Ch2-Laszlo3}. 
In the GBAR zone, the first formation of ortho-positronium with $\tau=142$ ns was observed in 2018. The expected density of ortho-positronium for $10^{10}$ positrons incident on the target is equal to $10^{11}$ ortho-Ps/cm$^3$ density in the cavity of size\break 1 mm x 1 mm x 20~mm.

According to calculations \cite{Ch2-Calculations} the cross-section for $\bar{H^{+}}$ formation can be enhanced for excited states of ortho-positronium. This hypothesis will be tested for the 3D state of ortho-positronium.

\section{Antihydrogen ion cooling and free fall experiment}

After the reaction, a beam composed of three species $\bar{p}$, $\bar{H}$ and $\bar{H}^{+}$ is electrostatically separated into three different directions, whereupon each beam is guided to a dedicated detection system.

The second part of the experiment is designed to catch and cool the antihydrogen ion \cite{Ch2-cooling1}. Once it reaches neV energy, the extra positron is photodetached, and the neutral, cold antihydrogen atom starts to fall.

\subsection{$\bar{H}^{+}$ capture and cooling}

The first step of the ultracold $\bar{H}^{+}$ preparation is to catch it in a cm scale RF linear trap. It is not possible to directly trap an ion of a few keV energy, this is why it is decelerated in the electrostatic field of a biased trap. Once the $\bar{H}^{+}$ energy is in the order of a few~eV it can be trapped. The energy dispersion (or spread) must be less than 20 eV in order to exceed 50\% trapping efficiency. The energy spread from the antiproton beam is transferred  to the anti ions. In ELENA this is of the order of 300~eV. Thus, an antiproton trap will be inserted in order to reduce this energy dispersion.

The antihydrogen ion does not have a transition allowing for direct laser cooling. It will be cooled by interaction with a cloud of simultaneously trapped and laser-cooled beryllium ions $Be^{+}$. In this procedure, the final temperature is higher than the Doppler limit (e.g. 0.47 mK or 60 neV in the case of $Be^{+}$), but can go down to 5-100 mK range. It is reached within a few ms \cite{Ch2-cooling2}. Once the antihydrogen ion is cold, it is injected to the precision RF trap where a single $Be^{+}$ ion is also trapped. In the precision trap the cold $Be^+$ and $\bar{H}^{+}$ ions are coupled harmonic oscillators on which ground state Raman sideband cooling can be performed \cite{Ch3-cooling3}. This method allows for the reduction of vibrational energetic levels. As a result, $\mu$K temperatures can be reached in less than 1 s. Both described cooling methods in dedicated traps were successfully tested by collaborators.

When the antihydrogen ion is cold, one of its two positrons can be photodetached by a $1.64~\mu$m laser. This moment defines the start of the free fall measurement.

\subsection{The free fall experiment}

The final free fall experiment takes place in the so-called ``free fall chamber''. It is a cylindrical chamber with at least $10^{-11}$ mbar pressure. The chamber is surrounded by detectors that are supposed to reconstruct the antihydrogen annihilation vertex:
\begin{itemize}
	\item tracker detectors - 50 cm by 50 cm MicroMegas chambers with a pion detection efficiency better than 96\% per plane \cite{Deb_thesis};
	\item 170 cm by 10cm by 5cm Time of Flight Detectors built from scintillator bars. The time resolution of these detectors is better than $80$ ps.
\end{itemize}

The first detector units were tested with cosmic radiation and antiproton annihilations.


\section*{Summary}

The GBAR experiment started to move to CERN in 2017. In 2018 the collaboration succeeded in developing the antihydrogen ion production part. The first tests with the antiproton beam supplied by the ELENA decelerator were performed. Currently, the proton source is under development in order to test the equipment with the symmetric hydrogen atom and ion production.

The positronium cloud preparation is progressing. The slow positron source reached\break $6\cdot10^{7}$~$e^{+}/s$ and the trapping is at the level of $10^{8}$ $e^{+}/$bunch. The first positronium cloud was made in the final reaction chamber. It is expected to produce hydrogen within a few months.

The cooling traps are tested by collaborators. The proof of cooling principle was done, however now the ion transport between two traps is being tested. The cooling efficiency is related to the energy dispersion of the antiproton beam. If it comes directly from ELENA it is too wide to be used in the experiment. However, this should be much improved with the insertion of an antiproton trap.

\section*{Acknowledgments}

We thank F. Butin and the EN team, C. Carli and the ELENA team as well as T. Eriksson and the AD
team for their fruitful collaboration.

\end{document}